\documentclass[10pt,a4paper]{article}
\usepackage{amsmath,amssymb,amsthm,graphicx,braket,multirow,authblk,amsthm,dcolumn,latexsym,subcaption,mathtools,cite,cancel,bm}
\usepackage[normalem]{ulem}
\usepackage[a4paper]{geometry}
\DeclareSymbolFontAlphabet{\amsmathbb}{AMSb}%

\usepackage[unicode=true,
bookmarks=true,bookmarksopen=false,
breaklinks=false,pdfborder={0 0 0},colorlinks=true]
{hyperref}
\usepackage{xcolor}

\definecolor{cblue}{rgb}{0.16, 0.32, 0.75}
\definecolor{cred}{rgb}{0.7, 0.11, 0.11}
\hypersetup{%
	,linkcolor=cred
	,citecolor=cblue
	,urlcolor=black
}
\def\<{\langle}
\def\>{\rangle}
\newtheorem{theorem}{Theorem}[section]
\newtheorem{proposition}{Proposition}[section]
\newtheorem{corollary}{Corollary}[section]

\newtheorem{remark}{Remark}
\newtheorem{Example}{Example}

\newcommand{\tr}{\mathop{\mathrm{Tr}}\nolimits}  \renewcommand{\Re}{\mathop{\mathrm{Re}}}
\def\oper{{\mathchoice{\rm 1\mskip-4mu l}{\rm 1\mskip-4mu l}
		{\rm 1\mskip-4.5mu l}{\rm 1\mskip-5mu l}}}

\renewcommand{\Re}{\mathop{\mathrm{Re}}}

\newcommand{\BH}{\mathcal{B}(\mathcal{H})}

\DeclareMathAlphabet\mathbfcal{OMS}{cmsy}{b}{n}

\begin{document}	
	\title{\textbf{Universal constraint for relaxation rates of semigroups of qubit Schwarz maps}}


	\author[$\hspace{0cm}$]{Dariusz Chru\'sci\'nski$^{1}$\footnote{darch@fizyka.umk.pl}, Gen Kimura$^2$\footnote{gen@shibaura-it.ac.jp}, and Farrukh Mukhamedov$^{3,4,5}$\footnote{farrukh.m@uaeu.ac.ae} }
	\affil[$1$]{\small Institute of Physics, Faculty of Physics, Astronomy and Informatics, Nicolaus Copernicus University, Grudziadzka 5/7, 87-100 Toru\'n, Poland}

\affil[$2$] {\small College of Systems Engineering and Science, Shibaura Institute of Technology, Saitama 330-8570, Japan}

\affil[$3$]{\small New Uzbekistan University, 54, Mustaqillik ave.,
 100007, Tashkent, Uzbekistan}

\affil[$4$]{\small School of Engineering, Central Asian University, 264, National Park Street,
Barkamol MFY, Yangiabad village, Qibray district,
111221, Tashkent region, Uzbekistan}

\affil[$5$]{\small Department of Mathematical Sciences, College of Science,
United Arab Emirates University,  P.O. Box 15551, Al Ain,  Abu Dhabi, UAE}
	
	\maketitle
	\vspace{-0.5cm}	
	
	\begin{abstract}
Unital qubit Schwarz maps interpolate between positive and completely positive maps. It is shown that relaxation rates of qubit semigroups of unital maps enjoying Schwarz property satisfy the universal constraint which provides a modification of the corresponding constraint known for completely positive semigroups. As an illustration we consider two paradigmatic qubit semigroups: Pauli dynamical maps and phase covariant dynamics. This result has two interesting implications: it provides a universal constraint for the spectra of qubit Schwarz maps and gives rise to a necessary condition for a Schwarz qubit map to be Markovian.

	\end{abstract}
	
	\maketitle
	
\section{Introduction}\label{sec:1}

Markovian dynamical semigroups are governed by the celebrated GKLS master equation \cite{GKS,L,Alicki} $\dot{\rho}_t = \mathcal{L}(\rho_t)$, with
\begin{equation}\label{GKLS}
  \mathcal{L}(\rho) = -i[H,\rho] + {\sum_{k}} \gamma_k \left( L_k \rho L_k^\dagger - \frac 12 \{ L_k^\dagger L_k,\rho\} \right) ,
\end{equation}
where $H$ denotes an effective system Hamiltonian, $L_k$ are noise operators, and $\gamma_k >0$ are positive transition rates (We set $\hbar=1$ {and} use the standard notations for the commutator and anti-commutator: $[A,B]:= AB - BA, \ \{A,B\}:= AB + BA$, respectively.)
It gives rise to {the general representation for the generator of} Markovian semigroup $\{\Lambda_t = e^{t \mathcal{L}}\}_{t\geq 0}$ of completely positive trace-preserving maps (CPTP) \cite{Paulsen,STORMER}. Solutions of GKLS master equation define very good approximations to evolutions of many real systems evolution, provided the system environment interaction is sufficiently weak and there is a large enough separation of timescales for the system and environment \cite{Alicki,Open1,Open2,Open3}.
The requirement for complete positivity gives rise to nontrivial constraints among relaxation rates which characterize the evolution of the system {\cite{GEN,KAW,PRL}}. {They} provide information on how fast the system relaxes to an asymptotic state and/or how fast it decoheres, {and are expected to serve as an experimental verification of the validity of complete positivity in open quantum dynamics.}
The spectrum of any GKLS generator is in general complex, however, if $\ell$ belongs to the spectrum then so does $\ell^*$. Moreover, there is a leading eigenvalue $\ell_0=0$ (the corresponding eigenvector corresponds to an invariant state of the evolution), and all remaining eigenvalues satisfy ${\rm Re}\, \ell_k \leq 0$. The corresponding relaxation rates are defined by $\Gamma_k := - {\rm Re}\, \ell_k$ (for more detailed exposition of spectral properties of maps and generators cf. e.g. \cite{WOLF,NM5,Amato}). It should be stressed that contrary to rates $\gamma_k$ which depend upon a particular representation of $\mathcal{L}$, the relaxation rates $\Gamma_k$  can be measured in the laboratory. The properties of relaxation properties of GKLS generators were studied by many authors
\cite{Alicki,Rel1,Rel2,Rel3,Rel4,Rel5}. For two-level systems a universal constraint was derived in \cite{GEN} where there are three
rates $\{\Gamma_1,\Gamma_2,\Gamma_3\}$ which satisfy
\begin{equation}\label{GEN}
  \Gamma_k \leq \frac 12 \Gamma , \ \ \ \Gamma := \Gamma_1+\Gamma_2+\Gamma_3 .
\end{equation}
{Interestingly, this reproduces \cite{GKS,GEN} the celebrated constraint between transversal $\Gamma_T$ and longitudinal $\Gamma_L$ rates:
\begin{equation}\label{2TL}
	2\Gamma_T \geq \Gamma_L .
\end{equation}
The above constraint was well tested in several experiments \cite{Alicki,Abragam}.
For any $d$-level system, universal constraints were derived in \cite{KAW,LAA}. Moreover, in \cite{PRL} it was conjectured that the following constraint is both the universal and the tight}
\begin{align}\label{UniCon}
	\Gamma_k \leq \frac 1d \Gamma,
\end{align}
where now $\Gamma= \Gamma_1+\ldots +\Gamma_{d^2-1}$. This provides generalization of (\ref{GEN}) for arbitrary (but finite) number of levels.

Here, `universality' means that the constraint is valid for any completely positive dynamical semigroup, i.e., for any GKLS master equation. The bound (\ref{UniCon}) is tight (that is, it cannot be further improved) since for any `$d$' there  exists a GKLS generator for which
the inequality  in \eqref{UniCon} is saturated.
This conjecture was verified for many well known examples including unital semigroups, so called Davies semigroups derived in the weak coupling limit and many others \cite{PRL} (cf. also \cite{LAA}).

In this paper we analyze the properties of qubit generators leading to semigroups of maps which are not necessarily completely positive but satisfy the Schwarz inequality in the (dual) Heisenberg picture (cf. next Section). All completely positive unital maps satisfy the Schwarz inequality but the converse is generally not true. However, all unital Schwarz maps are necessarily positive. Hence, in the qubit case, Schwarz maps interpolate between positive and completely positive unital maps. Schwarz maps are widely used in many aspects of mathematical physics (cf. the recent paper \cite{Alex}). In particular they are connected to several monotonicity properties (e.g. monotonicity of relative entropy) which are of great importance for quantum information theory  \cite{Petz}.

For semigroups of positive maps the corresponding generator can still be represented via formula (\ref{GKLS}) but some of the rates $\gamma_k$ can be negative. In this paper we prove (cf. Theorem \ref{MAIN}) that if $\mathcal{L}$ generates a semigroup of positive maps which satisfy (in the Heisenberg picture) the Schwarz inequality then the constraint (\ref{GEN}) is modified to
\begin{equation}\label{GEN1}
  \Gamma_k \leq \frac 2 3 \Gamma.
\end{equation}
It turns out that the above bound is also tight. It is, therefore, clear that for the qubit case one has
\begin{subequations}\label{GEN2}
\begin{equation}
  \Gamma_k \leq \frac 1 \alpha \Gamma ,
\end{equation}
where
\begin{equation}\label{}
  \alpha = \left\{ \begin{array}{ll} 1\ , & \ \ \mbox{for positive maps} \\ \frac 32\ , & \ \ \mbox{for Schwarz maps} \\ {2\ (=d)}. & \ \ \mbox{for completely positive maps} \end{array} \right.
\end{equation}
\end{subequations}
{In \cite{WOLF} (see also \cite{Alex}), Schwarz maps were referred to as $\frac{3}{2}$-positive maps, albeit as a `not too serious alternative'. It is intriguing that our results provide another evidence validating this nomenclature. In this paper we often use the term $\alpha$-positive map meaning: positive $(\alpha=1)$, Schwarz  $(\alpha=3/2)$, and completely positive $(\alpha=2)$ map. Note that,} the very condition (\ref{GEN1}) provides a modification of \eqref{2TL} as follows:
\begin{align}\label{4TL}
4 \Gamma_T \ge \Gamma_L,
\end{align}
which is exemplified by a phase-covariant qubit evolution (cf. Section \ref{S-III}).

The paper is organized as follows: Section \ref{S-II} provides a brief introduction to Schwarz maps and Markovian generators of semigroups of unital Schwarz maps. In Section \ref{S-III} we analyze two paradigmatic qubit semigroups of Pauli and phase-covariant maps. For these two examples necessary and sufficient for the corresponding generators to generate semigroups of Schwarz maps are derived. {These offer not only a simple illustration of the validity of {the constraint (\ref{GEN1})} but also provide examples that satisfy the equality, thereby demonstrating the `tightest' aspect of the constraint}. In section \ref{S-MAIN} we provide the proof of (\ref{GEN1}) for arbitrary qubit semigroup  enjoying the Schwarz property. As applications of our results  we discuss  in Section \ref{S-Spectra} necessary conditions for the spectra of unital $\alpha$-positive maps and
necessary conditions for $\alpha$-positive maps to be Markovian. 
Final conclusions are collected in Section \ref{S-C}. Additional technical details are presented in the Appendix.


\section{Preliminaries: Schwarz maps and Markovian generators}  \label{S-II}

Let $\mathcal{M}_n$ denote the space of $n \times n$ complex matrices. A unital linear map $\Phi : \mathcal{M}_n \to \mathcal{M}_n$, that is, $\Phi(\oper) = \oper$,  is called a Schwarz map if

\begin{equation}\label{KS}
  \Phi(X^\dagger X) \geq \Phi(X)^\dagger \Phi(X)  ,
\end{equation}
for all $X \in \mathcal{M}_n$ \cite{Paulsen,STORMER,KADISON,BHATIA}. It was shown by Kadison \cite{Kadison1,Kadison2} that if $\Phi$ is a positive unital map then it satisfies celebrated Kadison inequality

\begin{equation}\label{K}
  \Phi(X^2) \geq \Phi(X)^2   ,
\end{equation}
for all $X^\dagger=X$. However, not all positive unital maps satisfy (\ref{KS}). The simples example is provided by the transposition map.
Actually, Kadison inequality was generalized by Choi \cite{Choi1,Choi2} who proved that if $\Phi$ is positive and unital, then

\begin{equation}\label{K-choi}
  \Phi(X^\dagger X) \geq \Phi(X^\dagger) \Phi(X)  ,\ \ \  \Phi(X^\dagger X) \geq \Phi(X) \Phi(X^\dagger) ,
\end{equation}
for any normal operator $X$. Moreover, it was shown \cite{Choi1,Choi2} that if $\Phi(\oper) > 0$, then 2-positivity is equivalent to

\begin{equation}\label{K-choi-2}
  \Phi(X^\dagger T^{-1} X) \geq \Phi(X^\dagger) \Phi(T)^{-1} \Phi(X) ,
\end{equation}
for all $X,T\in \mathcal{M}_n$, such that $T$ is invertible. Actually, the above result maybe slightly generalized as follows \cite{Alex}: $\Phi$ is 2-positive iff   $\Phi(\oper) \geq 0$ and

\begin{equation}\label{Alex}
  \Phi(X^\dagger T^{-} X) \geq \Phi(X^\dagger) \Phi(T)^{-} \Phi(X) ,
\end{equation}
is valid for all $X \in \mathcal{M}_n$, and $T\geq 0$ with ${\rm ker}T \subseteq {\rm ker}X^\dagger$. In (\ref{Alex}) $T^-$ denotes the Moore-Penrose inverse \cite{GI2,GI3}.  It is, therefore, clear that if $\Phi(\oper)=\oper$, then (\ref{K-choi-2}) implies (\ref{KS}), i.e. any unital 2-positive map is a Schwarz map. {However, it is known that there exist  Schwarz maps which are not $2$-positive \cite{Choi1,Choi2} (cf. also \cite{WOLF}). Hence, for $n=2$ Schwarz maps interpolate between positive and completely positive maps unital maps.

Consider a dynamical semigroup $\{\Lambda_t\}_{t\geq 0}$ of linear positive trace-preserving maps on $\BH$, i.e. {for any $t,s \geq 0$ one has $\Lambda_{t+s} = \Lambda_t \circ \Lambda_s $,} $\Lambda_t(X)\geq 0$ for $X \geq 0$, and ${\rm Tr}\Lambda_t(X) = {\rm Tr}X$ for any $X \in \BH$. Passing to the dual Heisenberg picture via

\begin{equation}\label{}
  (\Lambda_t^\ddag(X),Y) := (X,\Lambda_t(Y)),
\end{equation}
where $(X,Y) = {\rm Tr}(X^\dagger Y)$ denotes the Hilbert-Schmidt inner product, one defines a semigroup $\{\Lambda^\ddag_t\}_{t\geq 0}$ of linear positive unital maps on $\BH$, i.e. $\Lambda^\ddag_t(\oper)=\oper$. A semigroup is uniquely defined by the corresponding generator $\mathcal{L} : \BH \to \BH$ via $\Lambda_t = e^{t \mathcal{L}}$. The map $\Lambda_t$ is trace-preserving if and only the corresponding generator $\mathcal{L}$ annihilates the trace ${\rm Tr}\,\mathcal{L}(X)=0$ for any $X \in \BH$. Equivalently, the dual generator annihilates identity operator in $\BH$, i.e.   $\mathcal{L}^\ddag(\oper) = 0$.

\begin{proposition}[\cite{Kos72}] $\Lambda_t=e^{t \mathcal{L}}$ is positive for all $t\geq 0$ if and only if

\begin{equation}\label{PQ}
 Q \mathcal{L}(P) Q \geq 0 ,
\end{equation}
for any pair of mutually orthogonal rank-1 projectors $P$ and $Q$.
\end{proposition}
Note, that (\ref{PQ}) is equivalent to  ${\rm Tr} [Q \mathcal{L}(P) Q] \geq 0$. Interestingly, the above property may be equivalently formulated as follows \cite{Evans-1979,Evans-1977} (cf. also \cite{NM5} for the recent review)

\begin{proposition} \label{P-KS} $\mathcal{L}^\ddag$ generates a semigroup $\{\Lambda_t^\ddag\}_{t\geq 0}$ of positive unital maps if and only if $\mathcal{L}^\ddag(\oper)=0$ and

\begin{equation}\label{KS-LH}
  \mathcal{L}^\ddag(X^2) \geq \mathcal{L}^\ddag(X) X + X \mathcal{L}^\ddag(X) ,
\end{equation}
for all $X =X^\dagger \in \mathcal{M}_n$.
\end{proposition}
Lindblad \cite{L} provided the following  condition for the generator $\mathcal{L}$ for which  $\{\Lambda^\ddag_t\}_{t\geq 0}  $ is a semigroup of Schwarz maps.

\begin{proposition}[\cite{L}]  \label{PRO-L} $\mathcal{L}^\ddag$ generates a semigroup $\{\Lambda_t^\ddag\}_{t\geq 0}$ of unital Schwarz maps if and only if $\mathcal{L}^\ddag(\oper)=0$ and

\begin{equation}\label{L!}
  \mathcal{L}^\ddag(X^\dagger X) \geq \mathcal{L}^\ddag(X^\dagger) X + X^\dagger \mathcal{L}^\ddag(X) ,
\end{equation}
for all $X  \in \mathcal{M}_n$.
\end{proposition}
It is evident from (\ref{L!}) that if both $\mathcal{L}^\ddag_1$ and $\mathcal{L}^\ddag_2$ generate Schwarz semigroups, then the sum $\mathcal{L}^\ddag_1 + \mathcal{L}^\ddag_2$ also generates a Schwarz semigroup.

Note that a generator of a unital Hermiticity-preserving semigroup has the following representation \cite{GKS}:
\begin{equation}\label{HPGen}
	\mathcal{L}^\ddag(X) = i[H,X] + \Phi^\ddag(X) - \frac 12 \{\Phi^\ddag(\oper),X\} ,
\end{equation}
{where $H = H^\dagger$ and $\Phi$ is a Hermiticity-preserving map. Using \eqref{L!}}, one finds \cite{L} that $\mathcal{L}^\ddag$ gives rise to a semigroup of Schwarz maps if and only if

\begin{equation}\label{Phi!}
  \Phi^\ddag(X^\dagger X) \geq \Phi^\ddag(X^\dagger) X + X^\dagger \Phi^\ddag(X) - X^\dagger {\Phi^\ddag(\oper)}X,
\end{equation}
for all $X  \in \mathcal{M}_n$. Actually, it is sufficient to check (\ref{Phi!}) for traceless operators. Indeed, {letting $X = X_0 + a \oper$, with traceless $X_0$, one} easily checks
\begin{equation}\label{}
  \Phi^\ddag(X^\dagger X) - \Phi^\ddag(X^\dagger) X - X^\dagger \Phi^\ddag(X) + X^\dagger {\Phi^\ddag(\oper)}X =  \Phi^\ddag(X_0^\dagger X_0) - \Phi^\ddag(X_0^\dagger) X_0 - X_0^\dagger \Phi^\ddag(X_0) + X_0^\dagger {\Phi^\ddag(\oper)}X_0 .
\end{equation}


\section{Paradigmatic qubit semigroups}   \label{S-III}

In this Section we analyze two paradigmatic qubit semigroups in terms of the corresponding relaxation rates $\Gamma_k$ ($k=1,2,3$).
Our analysis shows that the constraint (\ref{GEN2}) is satisfied.

\subsection{A semigroup of Pauli maps}

Consider the following generator
\begin{equation}\label{L-Pauli}
  \mathcal{L}(\rho) = \frac 12 \sum_{k=1}^3 \gamma_k (\sigma_k \rho \sigma_k - \rho),
\end{equation}
where $\gamma_k \ (k = 1,2,3)$ are real numbers and $\sigma_k \ (k=1,2,3)$ are the Pauli matrices (we use here the standard identification: $\sigma_1 = \sigma_x,\sigma_2 = \sigma_y,\sigma_3 = \sigma_z$). Note, that $\mathcal{L}^\ddag = \mathcal{L}$ and hence $\Lambda_t^\ddag = \Lambda_t$. Self-dual maps are necessarily unital.
One finds for the spectrum: $\mathcal{L}(\oper) = 0$, together with

\begin{equation}\label{}
  \mathcal{L}(\sigma_k) = \lambda_k \sigma_k ,
\end{equation}
with $\lambda_1 = -(\gamma_2+{\gamma_3})$, $\lambda_2 = -(\gamma_3+\gamma_1)$, and $\lambda_3 = -(\gamma_1+\gamma_2)$. Hence, the corresponding relaxation rates read

\begin{equation}\label{}
  \Gamma_1 = \gamma_2 + {\gamma_3} \ , \ \ \Gamma_2 = \gamma_3+\gamma_1 \ ,\ \ \Gamma_3 = \gamma_1+\gamma_2 .
\end{equation}
It is well known that (\ref{L-Pauli}) gives rise to a semigroup of positive maps if and only if all $\Gamma_k \geq 0$. Indeed, in terms of the corresponding Bloch vector ${\bm r} =(x_1,x_2,x_3)$, with $x_k =\tr(\rho\sigma_k)$, the evolution of a density operator corresponds to $x_k(t) = e^{-\Gamma_k t}$ and hence ${\bm r}(t)$ stays in Bloch ball if and only if $\Gamma_k \geq 0$. It shows that it is not necessary that all $\gamma_k$ are positive. Note, however, that at most only single $\gamma_k$ can be negative. Taking for example $\gamma_1=\gamma_2=\gamma$ and $\gamma_3=-\gamma$, one obtains $\Gamma_1=\Gamma_2=0$ and $\Gamma_3=2\gamma$.
Complete positivity is much more demanding and it requires all $\gamma_k \geq 0$ which is equivalent to the following relations between relaxation rates

\begin{equation}\label{GGG}
  \Gamma_i + \Gamma_j \geq \Gamma_k \ ,
\end{equation}
where $\{i,j,k\}$ are mutually different. The above relation can be compactly rewritten as follows

\begin{equation}\label{}
  \Gamma_k \leq \frac 12 \Gamma ,  \ \ \ k=1,2,3,
\end{equation}
where $\Gamma = \Gamma_1+\Gamma_2+\Gamma_3 = 2 (\gamma_1+\gamma_2+\gamma_3)$.

To check whether a positive map $\Lambda_t^\ddag$ satisfies Schwarz inequality one has to analyze (\ref{L!}). Assuming that $\gamma_1,\gamma_2\geq 0$, and taking $X=|0\>\<1|$ {(where $\ket{i} \ (i=0,1)$ denotes the normalized eigenvectors of $\sigma_3$, with eigenvalues $1,-1$, respectively)} one finds
\begin{equation}\label{}
\mathcal{L}^\ddag(X^\dagger X) - X^\dagger \mathcal{L}^\ddag(X) -  \mathcal{L}^\ddag(X^\dagger) X = {\frac{1}{2}} \begin{pmatrix}
    \gamma_1+\gamma_2 & 0 \\
    0 & \gamma_1+\gamma_2+4\gamma_3
  \end{pmatrix} ,
\end{equation}
which implies the following necessary condition
\begin{equation}\label{nS}
  \gamma_1+\gamma_2+4\gamma_3 \geq 0 .
\end{equation}
Observe, that in terms of relaxation rates condition (\ref{nS}) can be rewritten as follows
\begin{equation}\label{}
	\Gamma_1 + \Gamma_2 \geq \frac 12 \Gamma_3 ,
\end{equation}
which immediately implies $\Gamma \geq \frac 32 \Gamma_3$ and hence (\ref{GEN2}) holds since $\Gamma_3 = \max_k \Gamma_k$.

Interestingly, for a semigroup of Pauli maps the very constraint (\ref{GEN2}) may be equivalently rewritten as follows:
\begin{equation}\label{}
	\left\{ \begin{array}{ll} \Gamma_k \geq 0\ , & \ \ \mbox{for positive maps} \\ \Gamma_i + \Gamma_j \geq \frac 12 \Gamma_k \ , & \ \ \mbox{for Schwarz maps} \\ \Gamma_i + \Gamma_j \geq  \Gamma_k\ , & \ \ \mbox{for completely positive maps} \end{array} \right.
\end{equation}
where $\{i,j,k\}$ are all different.

To check that the bound (\ref{GEN2}) is tight we still need to construct a qubit Schwarz semigroup that saturates (\ref{GEN2}).
Recall, that \eqref{nS} is necessary but not sufficient for the Schwarz inequality to hold.
Assuming that $\gamma_3 < 0$ the following sufficient condition was derived in \cite{KS}
\begin{equation}\label{sS}
  \gamma_1 + 2\gamma_3 \geq 0 , \ \ \  \gamma_2 + 2\gamma_3 \geq 0 .
\end{equation}
(In the Appendix A we provide an independent proof of this result.)
Note, that when $\gamma_1=\gamma_2$ both conditions (\ref{nS}) and (\ref{sS}) coincide.
Hence, taking $\gamma_1 = \gamma_2 = -2 \gamma_3$ provides an example of a Schwarz semigroup that achieves the equality in (\ref{GEN2}).

\subsection{A semigroup of phase-covariant maps}

A linear map $\Phi : \mathcal{M}_2 \to \mathcal{M}_2$ is called to be a phase-covariant if

\begin{equation}\label{}
  U_\varphi \Phi(X) U^\dagger_\varphi = \Phi(  U_\varphi X U^\dagger_\varphi) ,
\end{equation}
where $U_\varphi = e^{-i \varphi \sigma_z}$, and $\varphi$ is an arbitrary (real) phase. The most general phase-covariant generator has the following form

\begin{equation}\label{1L+-}
  \mathcal{L}(\rho) = - i \frac{\omega}{2} [\sigma_z,\rho] + \gamma_+ \left(\sigma_+ \rho \sigma_- - \frac 12 \{ \sigma_-\sigma_+,\rho\}\right) + \gamma_-
  \left(\sigma_- \rho \sigma_+ - \frac 12 \{ \sigma_+\sigma_-,\rho\}\right) + \gamma_z (\sigma_z \rho \sigma_z - \rho) ,
\end{equation}
where {$\omega,\gamma_+,\gamma_-,\gamma_z$ are real numbers} and $\sigma_{\pm}=\frac{1}{2}(\sigma_{x}\pm i\sigma_{y})$. The above generator gives rise to a semigroup of CPTP maps if and only if $\gamma_\pm \geq 0$ and $\gamma_z \geq 0$. Let us analyze when the corresponding semigroup $\Lambda_t$ consists of positive maps and when $\Lambda_t^\ddag$ satisfies Schwarz inequality.

Note, that  $\rho_{\rm ss} = p_0 |0\>\<0| + p_1 |1\>\<1|$, with $p_0 = \gamma_+/(\gamma_+ +\gamma_-)$ and $p_1 = {\gamma_-}/(\gamma_+ + \gamma_-)$, defines a stationary state: $\mathcal{L}(\rho_{\rm ss})=0$. The spectral properties of $\mathcal{L}$ are characterized as follows:

\begin{equation}\label{}
  \mathcal{L}({\sigma_+}) = (- i \omega - \Gamma_T) {\sigma_+} \ , \ \ \   \mathcal{L}({\sigma_-}) = (i \omega - \Gamma_T) {\sigma_-} ,\ \ \
   \mathcal{L}(\sigma_z) = -\Gamma_L \sigma_z ,
\end{equation}
where the transversal $\Gamma_T$ and longitudinal $\Gamma_L$ relaxation rates are given by

\begin{equation}\label{GTGL}
  \Gamma_T = \frac{\gamma_+ + \gamma_-}{2} + 2 \gamma_z \ , \ \ \ \Gamma_L = \gamma_+ + \gamma_- .
\end{equation}
With the same reasoning as in Pauli semigroup, it is clear that positivity of $\Gamma_T$ and $\Gamma_L$ is necessary for positivity of $\Lambda_t$. However, contrary to the semigroups of Pauli maps this condition is not sufficient.

\begin{proposition}[\cite{Sergey}]\label{prop:Sergey} $\mathcal{L}$ defined by (\ref{1L+-}) gives rise to a semigroup of positive maps if and only if $\gamma_\pm \geq 0$ and
\begin{equation}\label{s2}
  \sqrt{\gamma_+ \gamma_-} + 2 \gamma_z \geq 0 .
\end{equation}
\end{proposition}
This result was already proved in \cite{Sergey} where the authors used quantum version of the Sinkhorn theorem.  Here we propose a simple proof based on the defining relation (\ref{PQ}). Let $P = |\psi\>\<\psi|$ and $Q= |\psi_\perp\>\<\psi_\perp|$, where $|\psi_\perp\>$ is a unique (up to a  phase factor) vector orthogonal to $|\psi\>$. Taking $\{|\psi\> = |0\>, |\psi_\perp\>=|1\>\}$ and $\{|\psi\> = |1\>, |\psi_\perp\>=|0\>\}$ one obtains
\begin{equation}\label{}
  {{\rm Tr}[Q \mathcal{L}(P) Q] = }  \langle 1 |  \mathcal{L}(|0\>\<0|) |1 \rangle =  \gamma_-  \geq 0 , \ \ \ \
  {{\rm Tr}[Q \mathcal{L}(P) Q] = }  \langle 0 |  \mathcal{L}(|1\>\<1|) |0 \rangle =  \gamma_+  \geq 0 .
\end{equation}
It is therefore clear that only $\gamma_z$ can be negative. Consider now {an arbitrary normalized vector $|\psi\> = z_1 |0\> + z_2|1\>$, with $|z_1|^2+|z_2|^2=1$}.
{Noting that $Q = \oper - |\psi\>\<\psi|$ and $\gamma_\pm \geq 0$, one has
\begin{eqnarray}\label{}
  && {\rm Tr}[Q \mathcal{L}(P) Q] = ( \sqrt{\gamma_-} |z_1|  - \sqrt{\gamma_+} |z_2|)^2 + 2(\sqrt{\gamma_+ \gamma_-} + 2 \gamma_z) |z_1| |z_2|
\end{eqnarray}
}
Now, it is evident that $ {{\rm Tr}[Q \mathcal{L}(P) Q]} \geq 0 $ if and only if $ \sqrt{\gamma_+ \gamma_-} + 2 \gamma_z \geq 0$, due to the fact that {$( \sqrt{\gamma_-} |z_1| - \sqrt{\gamma_+} |z_2|)^2$} is always non-negative.
\hfill $\Box$

In the Appendix {\ref{appB} we also show that $\gamma_\pm \geq 0$ and condition (\ref{s2}) are equivalent to
\begin{equation}\label{}
  \mathcal{L}^\ddag(X^2) - \{X,\mathcal{L}^\ddag(X)\} \geq 0 ,
\end{equation}
} for all $X^\dagger=X \in \mathcal{M}_2$, {which, combined with Proposition \ref{P-KS}, provides another proof of Proposition \ref{prop:Sergey}.}

Finally, the generator giving rise to a phase-covariant semigroup of positive maps can be characterized in terms of relaxation rates $\Gamma_T$ and $\Gamma_L$ and the parameter $\delta := \gamma_+ - \gamma_-$. One has
\begin{equation}\label{}
  \gamma_\pm = \frac 12 (\Gamma_L \pm \delta) , \ \ \ \gamma_z = \frac 14(2\Gamma_T - \Gamma_L) .
\end{equation}
and hence a phase covariant generator  $\mathcal{L}$ gives rise to semigroup of positive maps if and only if
\begin{equation}\label{I}
  \Gamma_L \geq |\delta| ,
\end{equation}
and
\begin{equation}\label{II}
  \sqrt{\Gamma_L^2 - \delta^2} + 2\Gamma_T - \Gamma_L \geq 0 .
\end{equation}

\begin{remark} Note, that for $\delta=0$ one has $\gamma_+=\gamma_-=\gamma$ and  the dissipative part of (\ref{1L+-}) reduces to the Pauli generator

\begin{equation*}\label{}
   \frac 12 \gamma (\sigma_x \rho \sigma_x + \sigma_y \rho \sigma_y - 2\rho) + \gamma_z (\sigma_z \rho \sigma_z - \rho) ,
\end{equation*}
with $\gamma_1=\gamma_2 = \gamma$ and $\gamma_3 = 2 \gamma_z$. In this case conditions (\ref{I}),(\ref{II})  reduce to $\Gamma_L \geq 0$ and $\Gamma_T\geq 0$.
\end{remark}

Consider now the dual generator

\begin{equation}\label{L+-}
  \mathcal{L}^\ddag(X) =  i \frac{\omega}{2} [\sigma_z,X] + \gamma_+ \left(\sigma_- X \sigma_+ - \frac 12 \{ \sigma_-\sigma_+,X\}\right) + \gamma_-
  \left(\sigma_+ X  \sigma_- - \frac 12 \{ \sigma_+\sigma_-,\rho\}\right) + \gamma_z (\sigma_z X \sigma_z - X).
\end{equation}

\begin{proposition}  The dual generator $\mathcal{L}^\ddag$ (see \eqref{L+-}) gives rise to a semigroup of Schwarz maps $\Lambda_t^\ddag$ if and only if $\gamma_\pm \geq 0$ together with

\begin{equation}\label{gg4}
  \gamma_\pm + 4 \gamma_z \geq 0 .
\end{equation}
\end{proposition}
Proof: taking $X=|0\>\<1|$ one finds

\begin{equation}\label{}
\mathcal{L}^\ddag(X^\dagger X) - X^\dagger \mathcal{L}^\ddag(X) -  \mathcal{L}^\ddag(X^\dagger) X =  {\begin{pmatrix}
    \gamma_+ & 0 \\
    0 & \gamma_+ +4\gamma_z
  \end{pmatrix} .}
\end{equation}
Similarly, for $X=|1\>\<0|$

\begin{equation}\label{}
\mathcal{L}^\ddag(X^\dagger X) - X^\dagger \mathcal{L}^\ddag(X) -  \mathcal{L}^\ddag(X^\dagger) X =  {\begin{pmatrix}
    \gamma_-+ 4 \gamma_z & 0 \\
    0 & \gamma_-
  \end{pmatrix} .}
\end{equation}
Hence, using Proposition \ref{PRO-L}, one finds that conditions $\gamma_\pm \ge 0$ and  (\ref{gg4}) are necessary in order to generate a semigroup of Schwarz maps. To show that they are also sufficient let $\gamma_z =: - \gamma$, with $\gamma > 0$. Clearly, if $\gamma_z \geq 0$, then $\mathcal{L}^\ddag$ generates completely positive unital and hence Schwarz semigroup. Introducing $\delta_\pm := \gamma_\pm - 4 \gamma$, one has
\begin{equation}\label{}
 { \gamma_\pm} \geq 0 , \ \ \  \delta_\pm \geq 0 .
\end{equation}
Let us observe that $\mathcal{L}^\ddag$ can be decomposed as $\,  \mathcal{L}^\ddagger = \mathcal{L}^\ddagger_1 + \mathcal{L}^\ddagger_2$, where

\begin{equation}\label{}
  \mathcal{L}^\ddagger_1(X) = {2\gamma (\sigma_x X \sigma_x + \sigma_y X \sigma_y - 2X) - \gamma(\sigma_z X \sigma_z - X)}
\end{equation}
and

\begin{equation}\label{}
  \mathcal{L}^\ddagger_2(X) =  i \frac{\omega}{2} [\sigma_z,X] + \delta_+ \left(\sigma_- X \sigma_+ - \frac 12 \{ \sigma_-\sigma_+,X\}\right) + \delta_-
  \left(\sigma_+ X  \sigma_- - \frac 12 \{ \sigma_+\sigma_-,X\}\right) .
\end{equation}
Now, {conditions \eqref{sS} implies that $\mathcal{L}^\ddagger_1$ generates a} semigroup of Pauli Schwarz maps {while} $\mathcal{L}^\ddagger_2$ generates a semigroup of completely positive and hence Schwarz maps. It is, therefore, clear that $\mathcal{L}^\ddagger_1 + \mathcal{L}^\ddagger_2$ generates a semigroup of Schwarz maps.



\begin{corollary} A phase covariant generator  $\mathcal{L}^\ddag$ gives rise to semigroup of Schwarz maps if and only if

\begin{equation}\label{III}
  \Gamma_L \geq |\delta| , \ \ \ \ \frac 12 (\Gamma_L - |\delta|) +  2\Gamma_T - \Gamma_L \geq 0 .
\end{equation}
\end{corollary}



\begin{corollary} If $\mathcal{L}^\ddag$ generates a semigroup of Schwarz maps, then

\begin{equation}\label{}
  \Gamma_T,\, \Gamma_L \leq \frac 23 \Gamma .
\end{equation}
\end{corollary}
Proof, indeed, conditions (\ref{gg4}) imply

\begin{eqnarray*}
  \frac 23 \Gamma = \frac 13 \Big( 4(\gamma_+ + \gamma_-) + 8 \gamma_z \Big)
  = \frac 13 \Big( 3(\gamma_+ + \gamma_-) + (\gamma_+ + 4\gamma_z) + (\gamma_- + 4\gamma_z)  \Big) \geq \gamma_+ + \gamma_- = \Gamma_L .
\end{eqnarray*}
{The proof for $\Gamma_T$ is straightforward due to positivity of $\Gamma_T$ and $\Gamma_L$.} Hence, it is shown that condition (\ref{GEN2}) holds for phase-covariant Schwarz semigroups.

\hfill $\Box$

{It is interesting to observe that the inequality $\Gamma_L \leq \frac 23 \Gamma$ results in a modification of the well-known relation \eqref{2TL} to \eqref{4TL} between longitudinal and transverse relaxation rates.} Recently, the qubit dynamical maps were analyzed in terms of the corresponding relaxation rates in \cite{PRA-23}. Both positivity and complete positivity is analyzed. Note, however, that authors of \cite{PRA-23} use different notation.

\section{General qubit semigroup of Schwarz maps}  \label{S-MAIN}

Any generator for a qubit trace-preserving semigroup can be represented in the basis of Pauli matrices as follows \cite{GKS,Alicki,Erika}

\begin{equation}\label{LC}
	\mathcal{L}(\rho) = - i[H,\rho] + {\sum_{i,j=1}^3} C_{ij} \left( \sigma_i \rho \sigma_j - \frac 12 \{\sigma_j \sigma_i,\rho\} \right) ,
\end{equation}
with Hermitian $3\times 3$ matrix $C_{ij}$, and Hermitian $H$. The evolution generated by (\ref{LC}) is completely positive if and only if $C_{ij}$ is positive definite. 
Observe, that the matrix $C$ can be decomposed as $C = S + i A$, where $S$ is real symmetric and $A$ is real antisymmetric. Now, let $\mathcal{O}$ be an orthogonal matrix which diagonalizes $S$, i.e. $S = \mathcal{O} D \mathcal{O}^T$, where $D$ is diagonal. Defining $\widetilde{C} := \mathcal{O}^T C \mathcal{O}$ one finds

\begin{equation}\label{}
	\widetilde{C} = D + i \mathcal{O}^T A \mathcal{O} ,
\end{equation}
which shows that  off-diagonal elements of $\widetilde{C}$ are purely imaginary, {as $\mathcal{O}^T A \mathcal{O} $ remains antisymmetric.}
Following \cite{GKS} it is convenient to parameterize $\widetilde{C}$ as follows

\begin{equation}\label{}
	D = g \oper - 2{\rm Diag}[g_1,g_2,g_3] , \ \ \ \mathcal{O}^T A \mathcal{O} = \left(
	\begin{array}{ccc}
		0 & -a_3 & a_2 \\
		a_3 & 0 & -a_1 \\
		-a_2 & a_1 & 0
	\end{array}\right) \ ,
\end{equation}
with {${\bm g} := (g_1,g_2,g_3), {\bm a} := (a_1,a_2,a_3) \in \mathbb{R}^3$ and $g = g_1+g_2+g_3$.  Note, that $g = {\rm Tr}\,C = {\rm Tr}\,\widetilde{C} = {\rm Tr}\,D$}.
Finally, one arrives at the following form

\begin{equation}\label{Cpara}
	\widetilde{C} = \left(
	\begin{array}{ccc}
		g_2+g_3 - g_1 & -ia_3 & ia_2 \\
		ia_3 & g_3 + g_1 - g_2 & -ia_1 \\
		-ia_2 & ia_1 & g_1+ g_2 -g_3
	\end{array}\right) .
\end{equation}
Now, defining  $F_a := \sum_i \mathcal{O}_{ia} \sigma_i$  one obtains

\begin{equation}\label{LC-2}
	\mathcal{L}(\rho) = - i[H,\rho] + {\sum_{i,j=1}^3} \widetilde{C}_{ij} \left( F_i \rho F_j - \frac 12 \{F_j F_i,\rho\} \right) .
\end{equation}
Note, that Hermitian matrices $F_i$ satisfies the same algebra as Pauli matrices, i.e. $F_k F_\ell = \delta_{k\ell} \oper + i \sum_m \epsilon_{k\ell m} F_m$. Let ${\bm r}=(x_1,x_2,x_3)$ be the corresponding Bloch representation of $\rho$ w.r.t. $F_i$ basis, that is,

\begin{equation}\label{}
	\rho = \frac 12 ( \oper + {\bm r}\cdot{\bm F} ) ,
\end{equation}
where ${\bm F}:=(F_1,F_2,F_3)$. The GKLS master equation $\dot{\rho}_t = \mathcal{L}(\rho_t)$ may be rewritten as the following equation for the evolution of the corresponding Bloch vector

\begin{align*}
	\frac{d}{dt} {\bm r}_t = - G {\bm r}_t + {\bm c}
\end{align*}
where ${\bm c} := 4{\bm a}$ and the matrix $G_{ij} := \frac 12 {\rm Tr}(F_i \mathcal{L}(F_j))$ reads
\begin{equation}\label{G}
	G = {2\left(
		\begin{array}{ccc}
			\medskip
			2g_1 &  h_3 & -h_2 \\
			\medskip
			-h_3 & 2g_2 & h_1 \\
			h_2 & -h_1 & 2g_3
		\end{array}
		\right) },
\end{equation}
where ${\bm h}:=(h_1,h_2,h_3)$ is defined via $H = \sum_k h_k F_k$. Note, that the spectrum of $\mathcal{L}$ apart from $\ell_0=0$ coincides with the spectrum of `$-G$', i.e. three relaxation rates $\Gamma_k$ are real parts of eigenvalues of $G$.

\begin{remark} For two paradigmatic qubit semigroups considered in the previous section we have $F_k=\sigma_k$. For a semigroup of Pauli maps one finds $g_k = {4 \Gamma_k}$ and clearly $h_k=0$ since $H=0$. For the phase-covariant case  one finds {$g_1=g_2=\Gamma_T/4$, $g_3=\Gamma_L/4$, and $h_3=\omega/2$}, that is,

	\begin{equation}\label{G2}
		G = {\left(
			\begin{array}{ccc}
				\medskip
				\Gamma_T &  \omega & 0 \\
				\medskip
				-\omega & \Gamma_T & 0 \\
				0 & 0 &  \Gamma_L
			\end{array}
			\right) }.
	\end{equation}
	The spectrum of $G$ reads: $\Gamma_L$, $\Gamma_T \pm i \omega$. Hence, for both semigroups the parameters $g_k$ {essentially} recover the corresponding relaxation rates. Note, however, that in general it is not true. It happens if and only if $G$ is diagonal, which corresponds to $H=0$, or there exists an axial symmetry: for example $g_1=g_2$ and $h_1=h_2=0$. These two cases correspond exactly to semigroups of Pauli maps and phase-covariant maps, respectively.
	
\end{remark}

\begin{proposition} \label{PRO} Let $G \in M_3(\mathbb{R})$ such that real parts of eigenvalues $\lambda_k \ (k=1,2,3)$ are all positive: $\Gamma_k := \Re \lambda_k \ge 0$. Then for any $\alpha \in [1,2]$
	\begin{align}\label{eq:1}
		\Gamma_k \le \frac{1}{\alpha } \tr G \ (k=1,2,3) \Longleftrightarrow f({\tr G}/{\alpha}) \ge 0
	\end{align}
	where $f(x) := \det (x \oper - G)$ is the characteristic polynomial {of $G$}.
	
\end{proposition}
Proof: the above result was already proved in \cite{GEN} for $\alpha=2$ corresponding to completely positive scenario. Here we provide a proof for any $\alpha\in [1,2]$. There are two separate cases: either exactly a single eigenvalue is real or all three are real.

\begin{itemize}

	\item Let the real eigenvalue be $\lambda_1 = \Gamma_R$, and others be complex conjugates, $\lambda_2 := \Gamma_C + i \Omega$ and $\lambda_3 := \Gamma_C - i \Omega$.
	The only nontrivial condition among three $\Gamma_k \leq  \frac{1}{\alpha } \tr G$ is
	\begin{align}\label{c0}
		\Gamma_R\le \frac{\tr G}{\alpha }.
	\end{align}
	The remaining condition $\Gamma_C \le \frac{\tr G}{\alpha }$ is trivially satisfied due $\alpha \le 2$. Actually, as $\tr G = \Gamma_R+ 2 \Gamma_C$, it follows that $\Gamma_C\le \frac{\tr G}{\alpha }$ if and only if $\Gamma_R + (2 -\alpha) \Gamma_C \geq 0$. Now, since $f(x) = 0$ has a single real root at $x = \Gamma_R$ and the function $f(x)$ satisfies $\lim_{x \to \pm \infty} f(x) = \pm \infty$, the very condition \eqref{c0} is equivalent to $f(\tr G/\alpha) \geq 0$.

	\item Let $\lambda_1 \ge \lambda_2 \ge \lambda_3 \ge 0$ be eigenvalues of $G$.  Then $\Gamma_k \leq  \frac{1}{\alpha } \tr G$ reduce to

	\begin{align}\label{c1}
		\lambda_1 \le \frac{1}{\alpha}\tr G.
	\end{align}
	Since $f(x) = (x-\lambda_1)(x -\lambda_2)(x-\lambda_3)$, one has that $f(x) \geq 0$ if and only if
	$$  \mbox{either}\ \lambda_2 \ge x \ge \lambda_3 , \ \ \ \mbox{or} \ \ x \ge \lambda_1 . $$
	Therefore, if condition \eqref{c1} holds, we have $f(\frac{\tr G}{\alpha }) \ge 0$. Conversely, if $f(\frac{\tr G}{\alpha }) \ge 0$, it follows that either \eqref{c1} or $\lambda_3 \le \frac{\tr G}{\alpha } \le \lambda_2$.
	However,  the latter case reduces again to \eqref{c1}.
	Indeed, if $\lambda_3 \le \frac{\tr A}{\alpha } \le \lambda_2$, one has $\lambda_1 + \lambda_3 \le (\alpha -1) \lambda_2$. As $\alpha \le 2$ and $\lambda_1 \ge \lambda_2 \ge \lambda_3 \ge 0$, this necessiates $\lambda_3 = 0$ and $\lambda_2 = \lambda_1$. However, in this case, \eqref{c1} is trivially satisfied as $\alpha \le 2$. (Note that in this case \eqref{c1} reads $\lambda_1 \le \frac{1}{\alpha} (\lambda_1 + \lambda_1 + 0) \Leftrightarrow 0 \le (2-\alpha)\lambda_1$.)
	
	\hfill $\Box$

\end{itemize}

\begin{theorem} \label{MAIN} If $\mathcal{L}^\ddag$ generates a semigroup of unital Schwarz maps, then $\Gamma_k \leq \frac 1\alpha \Gamma$ with $\alpha = 3/2$.
\end{theorem}
Proof: due to Proposition \ref{PRO} three conditions $\Gamma_k \leq \frac 1\alpha \Gamma$ are equivalent to $f(\tr G/\alpha) \geq 0$. One finds

\begin{eqnarray}
	{\frac{27}{2}} f(\frac{\tr G}{\alpha}) &=& {\Bigl(2 (g_2+g_3)-g_1\Bigr) \Bigl(2 \left(g_1+g_3\right)-g_2\Bigr) \Bigl(2 \left(g_1+g_2\right)-g_3\Bigr)}  \\
	&+& {9h_1^2\Bigl( 2 \left(g_2+g_3\right)-g_1\Bigr) + 9 h_2^2\Bigl(2 \left(g_1+g_3\right)-g_2\Bigr) + 9 h_3^2\Bigl(2 \left(g_1+g_2\right)-g_3\Bigr),} \nonumber
\end{eqnarray}
and hence to prove the Theorem it is enough to show
\begin{align}\label{gamma}
	2 (g_2+g_3)-g_1\ge 0,\  \ \ 2 \left(g_1+g_3\right)-g_2\ge 0,\  \ \ 2 \left(g_1+g_2\right)-g_3 \ge 0,
\end{align}
for any Schwarz semigroup. Now, due to Proposition \ref{PRO-L} $\mathcal{L}^\ddag$ has to satisfy the following inequality

\begin{equation}\label{}
	\mathcal{L}^\ddag(X^\dagger X) \geq \mathcal{L}^\ddag(X^\dagger) X + X^\dagger \mathcal{L}^\ddag(X) ,
\end{equation}
for all $X  \in \mathcal{M}_2$. Let $|f^{(k)}_a\>$ (for $a=0,1$) be eigenvectors of $F_k$, i.e.

\begin{equation}\label{}
	F_k = |f^{(k)}_0\>\< f^{(k)}_0| - |f^{(k)}_1\>\< f^{(k)}_1| .
\end{equation}
Note, that for $k \neq \ell$

\begin{equation}\label{}
	|\<  f^{(k)}_a| f^{(\ell)}_b\>|^2 = \frac 12 ,
\end{equation}
i.e. these three orthonormal basis are mutually unbiased. Using  $|f^{(3)}_a\>$ one finds for $X = |f^{(3)}_0\>\< f^{(3)}_1|$

\begin{equation}\label{}
	\mathcal{L}^\ddag(X^\dagger X) - \mathcal{L}^\ddag(X^\dagger) X - X^\dagger \mathcal{L}^\ddag(X) = \frac{1}{2}\left(
	\begin{array}{cc}
		a_3+g_3 & a_1-i a_2 \\
		a_1+i a_2 & -a_3+2 \left(g_1+g_2\right)-g_3 \\
	\end{array}
	\right),
\end{equation}
and

\begin{equation}\label{}
	\mathcal{L}^\ddag(XX^\dagger) - \mathcal{L}^\ddag(X) X^\dagger - X \mathcal{L}^\ddag(X^\dagger) = \frac{1}{2}\left(
	\begin{array}{cc}
		a_3+2 \left(g_1+g_2\right)-g_3 & a_1-i a_2 \\
		a_1+i a_2 & g_3-a_3 \\
	\end{array}
	\right).
\end{equation}
As the positivity of a matrix implies the positivity of its diagonal elements one finds

\begin{equation}\label{}
	2 (g_1+g_2)-g_3\ge |a_3| ,
\end{equation}
which implies desired inequality $ 2 (g_1+g_2)-g_3\ge 0$. Similarly, in the eigenbasis $|f^{(2)}_a\>$ of $F_2$ one obtains $ 2 (g_1+g_3)-g_2\ge 0$, and in the eigenbasis $|f^{(1)}_a\>$ of $F_1$ one obtains $ 2 (g_2+g_3)-g_1\ge 0$ which finally proves the Theorem. \hfill $\Box$

\section{Implications of the $\alpha$-bound: spectra and Markovianity}\label{S-Spectra}

Interestingly, the $\alpha$-bound  (\ref{GEN2}) implies nontrivial constraints for the spectra of unital $\alpha$-positive maps $\Phi^\ddag : \mathcal{M}_2 \to \mathcal{M}_2$. Let $\lambda_k \ (k=0,1,2,3)$ be eignevalues of $\Phi^\ddag$. By the unitality, one of the eigenvalues is $1$; we set $\lambda_0 = 1$ and $\lambda_k = {x_k} + i y_k \ (x_k,y_k \in \mathbb{R}, k=1,2,3)$ with the descending order $x_1\geq x_2 \geq x_3$.
Moreover, it is well known that the operator norm of any unital positive map is $1$ (see e.g. \cite{BHATIA}), hence $|\lambda_k| \le 1$ and also $|x_k| \le 1$.

{The key observation here is} that if $\Phi^\ddag$ is unital and $\alpha$-positive, with $\alpha \in \{1,\frac{3}{2},2\}$, then $\mathcal{L}^\ddag = \Phi^\ddag - {\rm id}$ {serves as} a generator of unital $\alpha$-positive maps. Indeed,
\begin{equation}\label{}
  \Lambda^\ddag_t = e^{t \mathcal{L}^\ddag} = e^{-t} e^{t \Phi^\ddag} = e^{-t} \Big( {\rm id} + t \Phi^\ddag + \frac{t^2}{2} \Phi^\ddag \circ \Phi^\ddag + \ldots \Big) .
\end{equation}
Now, taking into account that {positive linear} combination of $\alpha$-positive maps is $\alpha$-positive one immediately observes that $\Lambda_t$ is $\alpha$-positive. {Noting that the corresponding relaxation rates $\Gamma_k = 1-x_k$,} condition (\ref{GEN2}) implies
\begin{equation}\label{A}
	(\alpha-1)(1+x_3) \geq 2(\alpha-2) + x_1 + x_2.
\end{equation}
Now, for $\alpha=1$, condition (\ref{A}) is equivalent to $x_1+x_2 \leq 2$ which is trivially satisfied due to $|x_k| \leq 1$. For $\alpha =2$ it reduces to $1+x_3 \geq x_1 + x_2$, which in the case of Pauli maps recovers celebrated Fujiwara-Algoet condition \cite{FA}. Finally, for $\alpha=3/2$ it  implies $1+x_3 \geq 2(x_1+x_2-1)$.

\begin{Example} Consider a unitary map $\mathcal{U}(X) = UXU^\dagger$ with unitary operator $U= |0\>\<0|+ e^{i\phi} |1\>\<1|$ for some $\phi \in [0,2\pi)$. Clearly it is completely positive (and hence Schwarz).
The spectrum of $\mathcal{U}$ reads {$\{1,1,\exp(i\phi),\exp(-i\phi)\}$, hence $x_1 = 1, x_2 = x_3 = \cos \phi$}.
Condition (\ref{A}) is therefore  equivalent to `$1 \geq \cos\phi$' which is trivially satisfied.
\end{Example}

\begin{Example} Consider a transposition map $T: \mathcal{M}_2 \to \mathcal{M}_2$. One finds for the spectrum $\{1,1,1,-1\}$ which evidently violates (\ref{A}) {for both $\alpha = 3/2,2$ and hence one can conclude $T$ is neither completely positive nor Schwarz}.
Consider now the following deformation
\begin{equation}\label{Tp}
  T_p(X) = \frac{p}{2} \oper\, \tr X + (1-p) T(X) ,
\end{equation}
being a convex combination of $T$ and completely depolarizing map. {Note, that the spectrum of $T_p$ reads $\{1,\frac{1-p}{2},\frac{1-p}{2},-\frac{1-p}{2}\}$.
Thus, if the map \eqref{Tp} is Schwarz (resp. CP), condition (\ref{A}) implies that $p \ge 2/5$ (resp. $p \ge 2/3$). In \cite{OSID}, it was shown that the map $T_p$ is Schwarz for $p \geq \frac 12$ and completely positive for $p \geq \frac 23$. Hence, if $p\in [ \frac 25,\frac 12)$ the map $T_p$ satisfies (\ref{A}) but still it is not a Schwarz map, while the CP condition is entirely reproduced by (\ref{A}).}
\end{Example}

\begin{Example} A seminal reduction map $R: \mathcal{M}_2 \to \mathcal{M}_2$ is defined by

\begin{equation}\label{}
  R(X) = \oper\, \tr X - X ,
\end{equation}
and it is positive and unital \cite{Topical}. Its spectrum reads $\{1,-1,-1,-1\}$ and hence condition (\ref{A}) {(especially for $\alpha=2/3$)} is satisfied. However, $R$ is not a Schwarz map. Indeed, for $X = |0\>\<1|$ one finds

\begin{equation}\label{}
  R(XX^\dagger) = |1\>\<1| , \ \ \  R(X)R(X^\dagger) = |0\>\<0| ,
\end{equation}
and hence $R(XX^\dagger) \ngeq R(X)R(X^\dagger)$.   It clearly shows that (\ref{A}) is necessary but not sufficient condition for a unital map to satisfy Schwarz inequality.
\end{Example}

Given a unital $\alpha$-positive map $\Phi : \mathcal{M}_2 \to \mathcal{M}_2$ one may ask is it possible to find $\mathcal{L}$ such that $\Phi=e^\mathcal{L}$ (following \cite{MI,MII} one calls such map Markovian).  Note, that in this case eigenvalues $\lambda_k$ of $\Phi$ reads $\lambda_k = e^{\ell_k}$, where $\ell_k$ are eigenvalues of $\mathcal{L}$. Hence, if $\Gamma_k = - {\rm Re}\ell_k$ satisfy (\ref{GEN2}), then

\begin{equation}\label{}
	{\rm det}\, \Phi = \lambda_1 \lambda_2 \lambda_3 = e^{- \Gamma} \leq e^{-\alpha \Gamma_k} = |\lambda_k|^\alpha ,
\end{equation}
and hence

\begin{equation}\label{}
	({\rm det}\, \Phi)^{1/\alpha} \leq |\lambda_k| \leq 1 ,
\end{equation}
for any $k=1,2,3$. The above condition provides a universal constraint for the spectrum of qubit Markovian unital $\alpha$-positive map. In particular for the Pauli map one obtains

\begin{equation}\label{lll}
	\lambda_1\lambda_2 \leq \lambda_3^{\alpha-1} , \ \ \ \lambda_2\lambda_3 \leq \lambda_1^{\alpha-1} , \ \ \ \lambda_3\lambda_1 \leq \lambda_2^{\alpha-1} ,
\end{equation}
which for $\alpha=2$ (complete positivity) was independently derived in \cite{Mario,Karol,PRL}. For $\alpha=1$ it is trivially satisfied. However, for $\alpha=3/2$ it provides a necessary condition for the spectra of qubit Markovian unital Schwarz map:

\begin{equation}\label{lllS}
	\lambda_1\lambda_2 \leq \sqrt{\lambda_3} , \ \ \ \lambda_2\lambda_3 \leq \sqrt{\lambda_1} , \ \ \ \lambda_3\lambda_1 \leq \sqrt{\lambda_2} .
\end{equation}

\section{Conclusions}   \label{S-C}

In this paper we {have proved that relaxation rates for any qubit Schwarz semigroup satisfy the constraint (\ref{GEN1}), thereby completing the universal constraints \eqref{GEN2} for qubit semigroups with respect to positive, Schwarz, and completely positive maps.} This general result is illustrated by two paradigmatic qubit evolution: semigroups of Pauli maps and phase-covariant maps. For these semigroups it is simply possible to derive the conditions for the corresponding generator in terms of relaxation rates $\Gamma_k$ and then check validity of (\ref{GEN1}).
It should be stressed that the bounds (\ref{GEN2}) are tight. Indeed, consider a semigroup of Pauli maps: let $\gamma_1=\gamma_2=1$

\begin{itemize}
  \item  If  $\gamma_3=-1$, then $\Gamma_1=\Gamma_2=0$ and $\Gamma_3 =2$. Hence $\Gamma_3 = \Gamma$ saturates the bound for $\alpha=1$.
  \item If  $\gamma_3=-1/2$, then $\Gamma_1=\Gamma_2=1/2$ and $\Gamma_3 =2$. Hence $\Gamma_3 = \frac 23 \Gamma$ saturates the bound for $\alpha=3/2$.
   \item If $\gamma_3=0$, then $\Gamma_1=\Gamma_2=1$ and $\Gamma_3 =2$. Hence $\Gamma_3 = \frac 12 \Gamma$ saturates the bound for $\alpha=2$.
\end{itemize}
{We also presented two applications of the condition (\ref{GEN2}).
We derived a necessary condition \eqref{lll} for a trace preserving $\alpha$-positive map on qubit to be Markovian.
We also obtained a simple necessary condition \eqref{A} for a spectrum of unital $\alpha$-positive maps. }

Schwarz maps play important role in  mathematical physics \cite{Alex} and we hope that presented analysis contributes in a nontrivial way to the discussion on  the structure of Schwarz qubit maps.  Interestingly, our results  strongly support a proposal to call unital Schwarz map as $\frac 32$-positive (see also recent paper \cite{Alex}).

Finally, one may pose a natural question what happen to (\ref{GEN1}) beyond qubit scenario. Actually, we conjecture that the following constraint
 \begin{equation}\label{GENd}
  \Gamma_k \leq \frac 2 3 \sum_{\ell=1}^{d^2-1} \Gamma_\ell ,
\end{equation}
is satisfied for any semigroup of Schwarz maps of $d$-level quantum system. The clarification of this conjecture is postpone for the future research.

\appendix

\section{Appendix A}\label{appA}

In this appendix, we demonstrate that the conditions specified in \eqref{sS} provide sufficient criteria for semigroups of Schwarz Pauli maps:
\begin{proposition}\label{prop:Suf} If $\gamma_3 < 0$ and $\gamma_1,\gamma_2 \geq 2|\gamma_3|$, \eqref{L-Pauli} gives rise to a semigroup of Schwarz Pauli maps.
\end{proposition}
This was originally observed in \cite{KS}, but the proof presented here is independent. Indeed, we demonstrate it as a corollary of a more general result, Proposition \eqref{prop:Temp} below, which is of interest in its own right.

Let $\Delta$ be a dephasing channel w.r.t. eigenbasis of $\sigma_3$:
\begin{equation}\label{ap1}
\Delta(X) := P_0 X P_0 + P_1 X P_1 = \frac 12 (\sigma_z X \sigma_z + X)
\end{equation}
where $P_0 := |0\>\<0|= \frac{1}{2} (\oper + \sigma_z)$ and $P_1 :=|1\>\<1|= \frac{1}{2} (\oper - \sigma_z)$.
Consider a hermiticity-preserving map defined by
\begin{equation}\label{Psi}
	{\Psi_a(X)} := {\rm Tr}X \oper - a \Delta(X) .
\end{equation}
with a real parameter $a$. Note, that this map is self-dual $\Psi_a^\ddag = \Psi_a$ and $\Psi_a(\oper) = (2-a) \oper$. Define

\begin{equation}\label{}
  \mathcal{L}_a = \mathcal{L}^\ddag_a := \Psi_a - (2-a) {\rm id} .
\end{equation}

\begin{proposition}\label{prop:Temp} A linear map $\mathcal{L}_a$  generates a semigroup of  Schwarz maps if and only if $a \leq 3/2$.
\end{proposition}

Proof: {We need to show that the map \eqref{Psi} satisfies \eqref{Phi!} if and only if $a \leq 3/2$. We first show that necessarily $a \leq 3/2$. Indeed, defining the matrix $M_a$ by
\begin{equation}\label{}
	M_a := \Psi_a(X^\dagger X) - \Big(\Psi_a(X^\dagger) X + X^\dagger \Psi_a(X) - X^\dagger {\Psi_a(\oper)}X\Big),
\end{equation}
one finds $M_a = \oper + 2(1-a) |1\>\<1|$ for $X = |0\>\<1|$, and hence $M_a\geq 0$ iff $2(1-a) \geq -1$ which gives $a \leq 3/2$. To show that $a \leq 3/2$ is sufficient it is enough to consider traceless $X$.
Let
\begin{equation}\label{}
	X = \left( \begin{array}{cc} z & z_1 \\ z_2 & - z \end{array} \right) ,
\end{equation}
with $z,z_1,z_2 \in \mathbb{C}$. It is clear that one can freely multiply $X$ by a complex number and hence essentially there are two alternatives: $z=0$ or $z=1$. If $z=0$ one obtains
\begin{equation}\label{}
M_a =	\left(
	\begin{array}{cc}
		(3-2 a)|z_2|^2 + |z_1|^2   & 0 \\
		0 & (3-2 a) |z_1|^2 + |z_2|^2  \end{array}
	\right)
\end{equation}
which is positive semi-definite for all $z_1,z_2$ iff $a\leq 3/2$. If $z=1$ one obtains
\begin{equation}\label{}
	M_a = \left(
	\begin{array}{cc}
		(3-2 a)|z_2|^2  & 0 \\
		0 & (3-2 a) |z_1|^2 \end{array}
	\right) + N,
\end{equation}
where we define
$$
N = \left(
\begin{array}{cc}
	|z_1|^2 +4  & 2(z_1 - \overline{z_2}) \\
	2(\overline{z_1} - z_2) & |z_2|^2 + 4
\end{array}
\right).
$$
Observing that ${\rm Tr} N = 8 + |z_1|^2+|z_2|^2 \ge 0$ and
$$
{\rm det} N = |z_1|^2 |z_2|^2 - 8 \Re{z_1 z_2} + 16 \ge |z_1 z_2|^2 - 8 |z_1 z_2| + 16 = (4 + |z_1 z_2|)^2 \ge 0,
$$
$N$ is clearly positive semi-definite independent of $a$.
Therefore, $a \le 3/2$ is again sufficient for $M_a$ to be positive semi-definite.
\hfill $\Box$
}

\bigskip

Proposition \ref{prop:Suf} is easily obtained as a corollary of this proposition. Indeed, considering the worst case scenario of the conditions $\gamma_3 < 0$ and $\gamma_1,\gamma_2 \geq 2|\gamma_3|$, i.e. $\gamma_1=\gamma_2=\gamma$ and $\gamma_3 = - \frac 12 \gamma$, one finds for the Pauli generator \eqref{L-Pauli} that
\begin{eqnarray}\label{}
  \mathcal{L}(X) = \frac 12 \gamma \Big(\sigma_1 X \sigma_1 + \sigma_2 X \sigma_2 - \frac 12 \sigma_3 X \sigma_3 - \frac 32 X\Big) = \gamma\Big(  {\rm Tr}X \oper  - \frac{3}{2} \Delta(X) -  \frac{1}{2} X \Big),
\end{eqnarray}
where we have used the identity \cite{GKS}
\begin{equation}\label{ap2}
	\sum_{k=1}^3 \sigma_k X \sigma_k + X = 2 {\rm Tr}X \oper .
\end{equation}
In particular, for $a = \frac 32$,
\begin{equation}\label{}
  \mathcal{L}(X) = \gamma(\Psi_{\frac 32}(X) - \frac{1}{2} X).
\end{equation}
Noting $\Psi_{\frac 32}(\oper) = \oper/2$, this aligns with the representation (\ref{HPGen}) with $H = 0 $ and $\Phi(X) = \gamma \Psi_{\frac 32}(X)$.
Hence, Proposition \ref{prop:Temp} implies Proposition \ref{prop:Suf}, as intended.

\section{Appendix B}\label{appB}

We prove that the following condition {for a phase-covariant generator \eqref{L+-}:
\begin{equation}\label{LP}
  \mathcal{L}^\ddag(X^2) \geq \{\mathcal{L}^\ddag(X), X\},
\end{equation}
 for all $X =X^\dagger \in \mathcal{M}_2$ is equivalent to $\gamma_\pm \geq 0$ and condition (\ref{s2}), i.e., $\sqrt{\gamma_+ \gamma_-} + 2 \gamma_z \geq 0$.

{First, it is enough to check (\ref{LP}) for traceless operators.} Indeed, let $X = X_0 + a \oper$, with ${\rm Tr}\,X_0=0$. One has
\begin{equation}\label{1LP}
  \mathcal{L}^\ddag(X^2) - \{ X, \mathcal{L}^\ddag(X) \} =  \mathcal{L}^\ddag(X_0^2) - \{ X_0, \mathcal{L}^\ddag(X_0) \} ,
\end{equation}
due to $\mathcal{L}^\ddag(\oper)=0$. Moreover, due to the covariance property, it is sufficient to analyze
\begin{equation}\label{SufXZ}
		X = x \sigma_x + z \sigma_z \ , \ \ \ x,z \in \mathbb{R}.
\end{equation}
As $X^2 = (x^2+z^2) \oper$ and hence $\mathcal{L}^\ddag(X^2)=0$, one has
\begin{eqnarray*}
M &=& \mathcal{L}^\ddag(X^2) - \{\mathcal{L}^\ddag(X), X\} = - \{\mathcal{L}^\ddag(X), X\}  \\ &=& \left(
\begin{array}{cc}
	4 z^2 \gamma_+ + x^2 (\gamma_+ + \gamma_- + 4 \gamma_z) & 2 \left(\gamma _+-\gamma _-\right) x z \\
	2 \left(\gamma _+-\gamma _-\right) x z & 4 z^2 \gamma_- + x^2 (\gamma_+ + \gamma_- + 4 \gamma_z) \\
\end{array}
\right).
\end{eqnarray*}
Taking this into account, we first show that the positivity of $M$ implies $\gamma_\pm \geq 0$ and condition (\ref{s2}).
Indeed, taking $x=0,z=1$ (i.e., $X = \sigma_z$), one has

$$ M = \begin{pmatrix} 4 \gamma_+  & 0 \\ 0 &  4 \gamma_- \end{pmatrix} , $$
implying $\gamma_\pm \ge 0$. Incidentally, if we take $x = 1,z=0$ (i.e., $X = \sigma_z$), one has

$$ M = \begin{pmatrix} \gamma_+ + \gamma_- + 4 \gamma_z  & 0 \\ 0 &  \gamma_+ + \gamma_- + 4 \gamma_z \end{pmatrix} , $$
yielding
\begin{align}\label{eq:X=sigx}
\gamma_+ + \gamma_- + 4 \gamma_z \ge 0,
\end{align}
which is equivalent to the positivity of $\Gamma_T$ (cf. Eq.~\eqref{GTGL}). To get the condition (\ref{s2}), we compute the determinant of $M$ with $z =1$:
$$
{\rm det} M = a x^4 + b x^2 + c
$$
where

$$ a = 16 \gamma _- \gamma _+, \ \ \  b = 16 (\gamma _- \gamma _+ + \left(\gamma _-+\gamma _+\right) \gamma_z), \ \ \ c = ( \gamma_+ + \gamma_- + 4\gamma_z)^2 . $$
Note, that if $\gamma_+=0$ or $\gamma_-=0$, then $a=0$ and hence ${\rm det} M = b x^2 + c \geq 0$ only if $b\geq 0$ which implies $\gamma_z\geq 0$. Assuming $\gamma_\pm > 0$, one has ${\rm det} M \geq 0$  (for any $x \in \mathbb{R}$) if and only if the discriminant
\begin{align}\label{disc}
	b^2 - 4 ac = -64 \left(\gamma _- -\gamma _+\right){}^2 \left(\gamma _- \gamma _+-4 \gamma_z^2\right)
\end{align}
is non-positive.  Clearly, for $\gamma_+ \neq \gamma_-$ this implies  (\ref{s2}). If $\gamma_+ = \gamma_-$, then (\ref{s2}) is equivalent to (\ref{eq:X=sigx}).

{To show the sufficiency, assume $\gamma_\pm \ge 0$ and $\sqrt{\gamma_+ \gamma_-} + 2 \gamma_z \geq 0$.
As mentioned above, it is enough to show the positivity of $M$ for all $X$ of the form \eqref{SufXZ}. If $z =0$, one has 

$$ M  = \left(
\begin{array}{cc}
	\left(\gamma _-+\gamma _++4 \gamma _z\right) x^2 & 0 \\
	0 & \left(\gamma _-+\gamma _++4 \gamma _z\right) x^2 \\
\end{array}
\right) . $$
 However, this is shown to be positive through condition (\ref{s2}) by employing the Arithmetic Mean-Geometric Mean Inequality. In the case where $z \neq 0$, we may assume $z = 1$ without loss of generality. Now, the positivity of $\mathrm{Tr}M = 4 (\gamma_+ + \gamma_-) + 2 x^2 (\gamma_+ + \gamma_- + 4 \gamma_z)$ follows from the conditions (again using the AM-GM Inequality). Moreover, the positivity of $\mathrm{det} M$ also follows by observing that equation $b^2 - 4 ac$ in (\ref{disc}) is negative. Since $M \ge 0$ if and only if $\mathrm{Tr}M\geq 0$ and $ \mathrm{det}M \ge 0$, this completes the proof.
}

\section*{Acknowledgments}
 DC was supported by the Polish National Science Center project No. 2018/30/A/ST2/00837.


\begin{thebibliography}{1} \bibliographystyle{plain}

\bibitem{L} G. Lindblad, On the Generators of Quantum Dynamical Semigroups, Comm. Math. Phys. \textbf{48}, 119 (1976).

\bibitem{GKS} V. Gorini, A. Kossakowski, E.~C.~G. Sudarshan,
Completely positive dynamical semigroups of N-level systems, J. Math. Phys. \textbf{17}, 821 (1976).


\bibitem{Alicki} R. Alicki and K. Lendi, {\it Quantum Dynamical
Semigroups and Applications} (Springer, Berlin, 1987).



\bibitem{Paulsen} V. Paulsen, Completely Bounded Maps and Operator Algebras
(Cambridge University Press, Cambridge, UK, 2003).

\bibitem{STORMER} E. St{\o}rmer, Positive Linear Maps of Operator Algebras.
SpringerMonographs in Mathematics (Springer-Verlag, Berlin, 2013).

\bibitem{Open1} H.-P. Breuer and F. Petruccione, {\em The Theory of Open
Quantum Systems}, Oxford University Press, Oxford, 2007.


\bibitem{Open2} A. Rivas and S. F. Huelga, {\em Open Quantum Systems. An
Introduction} (Springer, Heidelberg, 2011).

\bibitem{Open3} C. Gardiner, P. Zoller, {\em Quantum Noise: A Handbook of Markovian and Non-Markovian Quantum Stochastic Methods
with Applications to Quantum Optics}, 4th Edition, Springer, Berlin, 2004.


\bibitem{GEN} G. Kimura, Restriction on relaxation times derived from the Lindblad-type master equations for two-level systems, Phys. Rev. A {\bf 66}, 062113 (2002).

\bibitem{KAW} {G. Kimura, S. Ajisaka, K. Watanabe, Universal Constraints on Relaxation Times for $d$-level GKLS master equations
, Open Syst. Inf. Dyn., 24, 1740009 (2017).}

\bibitem{PRL} D. Chru\'sci\'nski, G. Kimura, A. Kossakowski, and Y. Shishido, On the universal constraints for relaxation rates for quantum dynamical semigroup, Phys. Rev. Lett. {\bf 127}, 050401 (2021).

\bibitem{WOLF} M. Wolf, Quantum channels and operations: A guided tour, (2012). Lecture notes
available at https://www-m5.ma.tum.de/foswiki/pub/M5/Allgemeines/MichaelWolf/QChannelLecture.pdf

\bibitem{NM5} D. Chru\'sci\'nski, Dynamical maps beyond Markovian regime, Phys. Rep. {\bf 992}, 1-85 (2022).

\bibitem{Amato} D. Amato, P. Facchi, and A.  Konderak, Asymptotics of quantum channels, J. Phys. A: Math. Theor. {\bf 56}, 265304 (2023).

\bibitem{LAA}  D. Chru\'sci\'nski, R. Fujii, G. Kimura, and H. Ohno, Constraints for the spectra of generators of quantum dynamical semigroups, LIn. Alg. Appl. {\bf 630}, 293 (2021).

\bibitem{Rel1} K. Dietz, Decoherence by Lindblad motion, J. Phys. A {\bf 37}, 6143 (2004).

\bibitem{Rel2} B. Baumgartner, H. Narnhofer, and W. Thirring, Analysis of quantum semigroups with GKS-Lindblad generators: II. General, J. Phys. A: Math. Gen. {\bf 41}, 065201 (2008).

\bibitem{Rel3} S. G. Schirmer and A. I. Solomon, Constraints on relaxation rates for $N$-level quantum systems Phys. Rev. A {\bf 70}, 022107
(2004).

\bibitem{Rel4} P. R. Berman and R. C. O'Connell, Constraints on dephasing widths and shifts in three-level quantum systems, Phys. Rev. A {\bf 71}, 022501 (2005).

\bibitem{Rel5} D. K. L. Oi and S. G. Schirmer, Limits on the decay rate of quantum coherence and correlation, Phys. Rev. A {\bf 86}, 012121
(2012).


\bibitem{Alex} E. Carlen and A. M\"uller-Hermes, Characterizing Schwarz maps by tracial inequlities, Lett. Math. Phys. {\bf 113}, 17 (2023).

\bibitem{Petz} F. Hiai and D. Petz, From quasi-entropy to various quantum information quantities,
Publ. Res. Inst. Math. Sci., {\bf 48}, 525 (2012).




\bibitem{Abragam} A. Abragam, Principles of Nuclear Magnetism (Oxford
University Press, New York, 1961); C. P. Slichter, Principles
of Magnetic Resonance (Springer-Verlag, New York, 1990).












\bibitem{KADISON} R. V. Kadison and J. R. Ringrose, Fundamentals of the Theory of Operator Algebras. Graduate Studies in Mathematics Vol. 15
(Academic Press, New York, 1986).

\bibitem{BHATIA}  R. Bhatia, Positive Definite Matrices. Princeton Series in Applied Mathematics (Princeton University Press, Princeton, NJ,
2015).


\bibitem{Kadison1} R. V. Kadison, A generalized Schwarz inequality and algebraic invariants for $C^*$-algebras, Ann. Math. {\bf 56}, 494 (1952).

\bibitem{Kadison2} R. V. Kadison, On the orthogonalization of operator representations,
Amer. J. Math. 77 (1955), 600-620.

\bibitem{Choi1} M. D. Choi, A Schwarz inequality for positive linear maps on $C^*$-algebras,
Illinois J. Math. {\bf 18}, 565 (1974).

\bibitem{Choi2} M. D. Choi, Some assorted inequalities for positive linear maps on $C^*$-algebras,
J. Operator Theory {\bf 4}, 271 (1980).



\bibitem{GI2} A. Ben-Israel and N. E. Thomas, Generalized Inverses: Theory and Applications (Berlin: Springer, 2003).

\bibitem{GI3} R. A. Horn and C. R. Johnson, Matrix Analysis (Cambridge: Cambridge University Press, 1985).


\bibitem{Kos72} A. Kossakowski, On necessary and sufficient conditions for the generators
of a quantum dynamical semi-group, Bull. Acad. Polon. Sci., Ser. Sci.
Math. Astronom. Phys. {\bf 20,} 1021 (1972).


\bibitem{Evans-1979} D. E. Evans and H. Hanche-Olsen, The generator of positive semigroups, J. Func. Anal. {\bf 32}, 207 (1979).


\bibitem{Evans-1977} D. E. Evans, Conditionally Completely Positive Maps on Operator Algebras,  Quart J. Math. Oxford {\bf 28}, 369 (1977).


\bibitem{KS} D. Chru\'sci\'nski and  F. Mukhamedov,  Dissipative generators, divisible dynamical maps, and the Kadison-Schwarz inequality, Phys. Rev. A. {\bf 100},  052120 (2019).

\bibitem{Sergey} S. N. Filippov, A. N. Glinov, and L. Lepp\"aj\"arvi, Phase covariant qubit dynamics and divisibility,  Lobachevskii J. Math. {\bf 41}, 617-630 (2020). (available as arXiv:1911.09468).

\bibitem{Erika} M. J. W. Hall, J. D. Cresser, Li. Li, and E. Andersson, Canonical form of master equations and characterization of non-Markovianity, Phys. Rev. A {\bf 89}, 042120 (2014).



\bibitem{PRA-23} G. Th\'eret and D. Sugny, Complete positivity, positivity, and long-time asymptotic behavior in a two-level open quantum system
Phys. Rev. A {\bf 108}, 032212 (2023).

\bibitem{FA} A. Fujiwara and P. Algoet, One-to-one parametrization of quantum channels, Phys. Rev. A {\bf 59}, 3290
(1999).


\bibitem{OSID} D. Chru\'sci\'nski and  F. Mukhamedov, On Kadison-Schwarz Approximation to Positive Maps, Open Sys. Inf. Dyn. {\bf 27}, 2050016 (2020).

\bibitem{Topical}  D. Chru\'sci\'nski and G. Sarbicki,  Entanglement witnesses: construction, analysis and classification, J. Phys. A {\bf 47},  483001 (2014).

\bibitem{MI} M. M. Wolf and I. Cirac, Dividing Quantum Channels,  Commun. Math. Phys. {\bf 279}, 147
(2008).

\bibitem{MII} M. M. Wolf, J. Eisert, T. S. Cubitt, and J. I. Cirac, Assessing Non-Markovian Quantum Dynamics, Phys. Rev.
Lett. {\bf  101}, 150402 (2008).


\bibitem{Mario} D. Davalos, M. Ziman, and C. Pineda, Divisibility of qubit channels and dynamical maps, Quantum {\bf 3}, 144
(2019).

\bibitem{Karol}  Z. Pucha{\l}a, L. Rudnicki, and K. \.Zyczkowski, Pauli semigroups and unistochastic quantum channels,  Phys. Lett. A
{\bf 383}, 2376 (2019).

\end{thebibliography}
\end{document}